\documentclass[12pt]{article}
\usepackage{amsmath,amssymb}
\usepackage{epsfig} 
\usepackage{color}\def\m@th{\mathsurround=0pt }
\def\leftrightarrowfill{$\m@th \mathord\leftarrow \mkern-6mu
        \cleaders\hbox{$\mkern-2mu \mathord- \mkern-2mu$}\hfill
        \mkern-6mu \mathord\rightarrow$}

\def\overleftrightarrow#1{\vbox{\ialign{##\crcr
        \leftrightarrowfill\crcr\noalign{\kern-1pt\nointerlineskip}
        $\hfil\displaystyle{#1}\hfil$\crcr}}}

\newcommand{\be}{\begin{equation}}
\newcommand{\ee}{\end{equation}}

\newcommand{\TeV}{\,\mathrm{TeV}}
\newcommand{\GeV}{\,\mathrm{GeV}}
\newcommand{\MeV}{\,\mathrm{MeV}}

\newcommand{\eV}{\,\mathrm{eV}}

\def\beq{\begin{equation}}
\def\eeq{\end{equation}}
\def\bea{\begin{eqnarray}}
\def\eea{\end{eqnarray}}
\begin{document}

\baselineskip=18pt

\setcounter{footnote}{0}
\setcounter{figure}{0}
\setcounter{table}{0}

\begin{titlepage}

\vspace{.3in}

\begin{center}

{\Large \bf Low-Scale Leptogenesis and the Domain Wall Problem in 
Models with Discrete Flavor Symmetries}

\vspace{0.5cm}

 {\bf Francesco Riva\footnote{email: riva$@$pd.infn.it}}
\vskip 0.5cm

\centerline{{\it Dipartimento di Fisica, Universit\`a di Padova, Italy}}

\end{center}
\vspace{.8cm}

\begin{abstract}
\medskip
\noindent
We propose a new mechanism for leptogenesis, which is naturally realized in models with a flavor symmetry based on the discrete group $A_4$, where the symmetry breaking parameter also controls the Majorana masses for the heavy right handed (RH) neutrinos. During the early universe, for $T\gtrsim \TeV$, part of the symmetry is restored, due to finite temperature contributions, and the RH neutrinos remain massless and can be produced in thermal equilibrium even at temperatures well below the most conservative gravitino bounds. Below this temperature the phase transition occurs and they become massive, decaying out of equilibrium and producing the necessary lepton asymmetry. Unless the symmetry is broken explicitly by Planck-suppressed terms, the domain walls generated by the symmetry breaking survive till the quark-hadron phase transition, where they disappear due to a small energy splitting between the $A_4$ vacua caused by the QCD anomaly.

\end{abstract}

\bigskip
\bigskip

\end{titlepage}


\section{Introduction}

Leptogenesis provides a natural scenario to explain the observed baryon abundance, where an asymmetry in the leptonic sector is initially generated by the out-of-equilibrium decay of heavy right-handed (RH) neutrinos and then distributed into baryons by so-called sphaleron processes \cite{LG}. The most attractive feature of leptogenesis is that it relies uniquely on the addition to the Standard Model of three RH neutrinos, addition which is already very welcome to explain the smallness of light neutrino masses via the see-saw mechanism \cite{seesaw}. This synergy between neutrino physics and the origin of matter underlines the need for studying leptogenesis in models with realistic neutrino mixing.

While our understanding of the origin of fermion masses and mixing angles remains at a primitive level, flavor symmetries provide a practical mean to replicate, within experimental uncertainties, the observed data. As originally proposed by Froggat and Nielsen \cite{Froggatt:1978nt}, a broken $U(1)$ flavor symmetry can be responsible for the small ratios between masses in the quark sector. Similarly, tri-bimaximal neutrino mixing, in excellent agreement with observations from neutrino oscillation experiments, can be reproduced within models with broken discrete non-abelian symmetries in the leptonic sector where sub-leading corrections in the symmetry breaking parameter account for small deviations from exact tri-bimaximal mixing \cite{Altarelli:2005yp}.

In this article, 
we study leptogenesis in models with natural tri-bimaximal mixing from a flavor symmetry based on the discrete group $A_4$ \cite{Altarelli:2005yx}. The arguments we use, however, can be easily extended to a more general class of models where the scale of lepton number violation is introduced by a spontaneous symmetry breaking mechanism. Symmetries that are spontaneously broken today might be restored during the early universe, due to finite temperature effects. We find that this symmetry restoration provides a natural scenario for thermal leptogenesis to work efficiently even at low reheat temperatures $T_{RH}\lesssim 10^{4\div5}\GeV$, well below the most stringent bounds from gravitino overproduction, with no need for the addition of  structure or fine-tuning. Indeed, a crucial aspect of the $A_4$-based models  is the structure of the breaking sector and the particular vacuum alignment that reproduces tri-bimaximal mixing in the neutrino sector. The symmetry is partially broken (down to a subgroup isomorphic to $Z_2$) by the vacuum expectation value (VEV) of a flat direction that obtains a potential  only in the presence of soft supersymmetry (SUSY) breaking terms. During the early universe, the finite energy density of the inflationary vacuum or the different occupation numbers of fermions and bosons in the thermal bath after reheating, break SUSY and contribute a (positive\footnote{For particular forms of the K\"ahler potential is it also possible to arrange for a negative mass of the flat-direction during inflation. In this case, the peculiar dynamics of flat directions in the early universe \cite{Basboll:2007vt,Giudice:2008gu} can reproduce the observed baryon asymmetry in a non-thermal way, as in \cite{Giudice:2008gu} or via Affleck-Dine baryogenesis \cite{Affleck:1984fy}.}) mass-squared term for flat directions \cite{Dine:1995uk}, restoring part of the flavor symmetry. The relevant aspect for leptogenesis is that, in these models, the lepton number violating Majorana masses of RH neutrinos are also proportional to the flavor symmetry breaking parameter and therefore vanish in the early universe. Thus, at early times, the would-be heavy RH neutrinos are effectively massless states and remain in equilibrium with the thermal bath down to temperatures of order the SUSY breaking soft masses, $\tilde{m}\sim {\cal O}(10^2-10^3)~\GeV$. Below this temperature, the thermal corrections to the flat direction potential become too small and its origin unstable: a smooth phase transition takes place in which the flat direction VEV rolls to large values, breaking the flavor group and lepton number and giving large Majorana masses $M_i$ to the RH neutrinos, which suddenly find themselves out of equilibrium at $T\ll M_i$. This is the perfect starting point for standard thermal Leptogenesis: an equilibrium abundance of RH neutrinos, ready to decay via CP-violating interactions producing a large lepton asymmetry. At temperatures above $\sim 100~\GeV$, sphaleron processes convert this asymmetry into the observed baryon asymmetry \cite{kuzmin}.

A potential problem of discrete symmetries broken spontaneously at low temperatures is the creation of domain walls. Their energy density red-shifts slower than that of matter or radiation and eventually, independently of their initial abundance, they come to dominate the energy density of the whole universe, leading to cosmological consequences incompatible with observations, such as imprinting large signatures in the cosmic microwave background \cite{DW}. We will show that, in the $A_4$-based models, the putative discrete symmetry is not one in fact: it is broken explicitly by the QCD anomaly that lifts the degeneracy between vacua separated by domain wall and drives them to collapse. Thus, unless the $A_4$ symmetry is also broken explicitly by the gravitational interactions (in which case the domain walls do not form at all), the domain wall network disappears at the QCD scale and standard cosmology is recovered.

In the next section we review models based on the $A_4$ flavor symmetry to reproduce tri-bimaximal mixing,  paying particular attention to the structure of the symmetry breaking sector. In section \ref{SecThermal} we discuss the dynamics of the flavor symmetry breaking during the early universe and the thermal production of RH neutrinos. In section \ref{LG} and \ref{secDW} we conclude with a study of leptogenesis and a note on the domain wall problem.


\section{Tri-bimaximal Neutrino Mixing from a Discrete Symmetry, at $T=0$}

Fits to neutrino oscillation data \cite{Schwetz:2008er} give the following values for the mixing angles,
\begin{equation}\label{data}
\sin^2\theta_{13}=0.01^{+0.016}_{-0.011}, \quad
\sin^2\theta_{23}=0.50^{+0.07}_{-0.06},\quad
\sin^2\theta_{12}=0.304^{+0.022}_{-0.016},
\end{equation}
which is compatible with tri-bimaximal mixing \cite{Harrison:2002er},
\begin{equation}\label{TB}
\sin^2\theta_{13}=0,\quad
\sin^2\theta_{23}=1/2,\quad
\sin^2\theta_{12}=1/3.
\end{equation}
It is tempting to explain this peculiar mixing pattern in terms of symmetries. In \cite{Altarelli:2005yx}, a model based on the discrete group $A_4$ was proposed, providing a dynamical mechanism which naturally reproduces this particular mixing matrix. A crucial role in this picture is played by the sector responsible for breaking the $A_4$ symmetry: only for a particular vacuum structure can the tri-bimaximal form of the mixing matrix be reproduced. In what follows we will review how this works.

The full flavor symmetry group is $A_4\times U(1)_{FN}\times Z_3$, where the $A_4$ part gives the required form of the mixing matrix, while the $U(1)_{FN}$ Froggatt-Nielsen factor provides the usual hierarchy among Yukawas for the charged leptons, through insertions of VEVs of the flavon fields $\Theta$, $\tilde{\Theta}$ suppressed by some unspecified large scale $\hat{\Lambda}$. The extra $Z_3$ guarantees that charged leptons and neutrinos couple to two different set of fields at leading order. Thank to this separation, the $A_4$ symmetry can be broken down to two different subgroups in the neutrino and charged lepton sectors; the mismatch between the tw different residual symmetries is what gives origin to the mxing. The flavor symmetry breaking sector includes two $A_4$ triplets, $\varphi^S$ and $\varphi^T$ and two singlets, $\xi$ and $\tilde{\xi}$ which couple directly to the lepton sector, but also a set of "driving fields", $\varphi_0^S$, $\varphi_0^T$ and $\xi_0$ that are needed to build a non-trivial scalar potential in the symmetry breaking sector. Beside the usual field content of the MSSM, there are also three RH neutrino fields, in a triplet of $A_4$. The charge assignments are as follows\footnote{The $U(1)_{FN}$ and $Z_3$ charge assignments given here are different from those of references \cite{Altarelli:2005yx,Altarelli:2005yp}. This choices doesn't alter their conclusion, but will be relevant for our discussion of leptogenesis in section \ref{LG} and domain walls in section \ref{secDW}. Note furthermore that, while the $A_4$ factor is crucial to obtain tri-bimaximal mixing, the $Z_3$ can be substituted by any (larger) symmetry that keeps the charged and neutrino sector separated.} (see \cite{Altarelli:2005yx} for a review on the $A_4$ group) :
\begin{center}
\begin{tabular}{|c||c|c|c|c|c||c|c|c|c|c||c|c|c||c|c|}
\hline
{\tt Field}& l & $e^c$ & $\mu^c$ & $\tau^c$ &$N^c $& $h_{u,d}$ & 
$\varphi^T$ & $\varphi^S$ & $\xi$ & $\tilde{\xi}$ & $\varphi_0^T$ & $\varphi_0^S$ & $\xi_0$&$\Theta$&$\tilde{\Theta}$\\
\hline
$A_4$ & $3$ & $1$ & $1'$ & $1''$ &$3$& $1$ & 
$3$ & $3$ & $1$ & $1$ & $3$ & $3$ & $1$&$1$&$1$\\
\hline
$Z_3$ & $\omega$ & $\omega$ & $\omega$ & $\omega$ &$\omega$ & $\omega$ &
$1$ & $\omega$ & $\omega$ & $\omega$ & $1$ & $\omega$ & $\omega$&$1$&$1$\\
\hline
$U(1)_R$ & $1$ & $1$ & $1$ & $1$ &$1$ & $0$ & 
$0$ & $0$ & $0$ & $0$ & $2$ & $2$ & $2$&$0$&$0$\\
\hline
$U(1)_{FN}$ & $-n$ & $2+n$ & $1+n$ & $n$ &$0$ & $0$ & 
$0$ & $0$ & $0$ & $0$ & $0$ & $0$ & $0$&$-1$&$1$\\
\hline
\end{tabular}
\end{center}
where included is also a $U(1)_R$ symmetry, containing the usual R-Parity of SUSY, that differentiates between matter fields, symmetry breaking sector fields and driving fields. The superpotential can be divided in two parts,
\begin{equation}
W=W_l+W_d.
\end{equation}
The lepton part includes, at leading order, the terms responsible for lepton masses and is given by (for details on how the $A_4$ indices are contracted, see \cite{Altarelli:2005yx})
\bea
W_l&=&\tilde{y}_e\; t^2 e^c (\frac{\varphi^T}{\Lambda} l)H_d+\tilde{y}_\mu \;t \;\mu^c (\frac{\varphi^T}{\Lambda} l)'H_d\nonumber\\
&+& y_\tau \tau^c (\frac{\varphi^T}{\Lambda} l)''H_d+ \tilde{y}\; t^n (N^c  l)H_u+
(x_A\xi+\tilde{x}_A\tilde{\xi}) (N^c N^c )+x_B (\varphi^S N^c N^c )+...
\label{wlss}
\eea
where $\Lambda$ is the cut-off of the theory and $t=\langle\tilde{\Theta}\rangle/\hat{\Lambda}\approx\langle\Theta\rangle/\hat{\Lambda}$ is the Froggatt-Nielsen factor that reproduces the mass hierarchy in the charged lepton sector. We shall absorbe the factors of $t$ in a redefinition of the yukawas,
\beq
\tilde{y}_e t^2= y_e,\quad \tilde{y}_\mu t= y_\mu, \quad \tilde{y} t^n= y,
\eeq
which allows us to treat $y$ as a free parameter in the following discussion. The driving part, which produces the right vacuum alignment, is
\bea
\label{wd}
W_d&=&a M (\varphi_0^T \varphi^T)+ g_0 (\varphi_0^T \varphi^T\varphi^T)\nonumber\\
&+&g_1 (\varphi_0^S \varphi^S\varphi^S)+
g_2 \tilde{\xi} (\varphi_0^S \varphi^S)+
g_3 \xi_0 (\varphi^S\varphi^S)+
g_4 \xi_0 \xi^2+
g_5 \xi_0 \xi \tilde{\xi}+
g_6 \xi_0 \tilde{\xi}^2+...
\eea
Here the coupling constants $x_{A,B}$ are generally complex, while the Yukawas $y$,$y_{e,\mu,\tau}$ can always be made real by a redefinition of the fields; dots stand for higher order operators in the $1/\Lambda$ expansion. The mass-parameter $M$  is taken to originate microscopically. A crucial aspect of this model is the existence of a flat direction at all orders in perturbation theory (in the limit of exact SUSY). This can be seen as follows. Since the fields in the flavor symmetry breaking and driving sectors are singlets under the gauge group, in the supersymmetric limit the scalar potential involves only F-terms (assuming canonical K\"ahler terms),
\begin{equation}\label{VSUSY}
V=\sum_i\left |\frac{\partial W}{\partial \phi_i}\right |^2,\quad \textrm{with}\quad \phi_i=\varphi^{T,S}_{(0)},\xi_{(0)}, \tilde{\xi}.
\end{equation}
This means that the conditions for the minimum of the potential are simply
\beq\label{MinimumCond1}
 \frac{\partial W}{\partial\varphi_{0,i}^{T,S}}=\frac{\partial W}{\partial\xi_{0}}=0,\quad i=1,2,3
 \eeq
 and
 \beq\label{MinimumCond2}
 \frac{\partial W}{\partial\varphi_{i}^{T,S}}=\frac{\partial W}{\partial\tilde{\xi}}=\frac{\partial W}{\partial\xi}=0,\quad i=1,2,3,
\eeq
where the subscript $i$ indicates the $A_4$-component of the triplets $\varphi^{T,S}_{(0)}$. Furthermore, since the driving fields carry two units of $R$-charge, the superpotential must be linear in these fields (see eq.(\ref{wd}), at the renormalizable level). This means that the 8 equations with 7 variables expressed in eq.(\ref{MinimumCond2}) always have the trivial solution $\langle \varphi_{0,i}^{T,S}\rangle=\langle \xi_{0}\rangle=0$. On the other hand, the remaining 7 equations in 8 variables expressed in eq (\ref{MinimumCond1}) do not contain the driving fields and do not exhibit a unique solution, but rather a flat direction of solutions, undetermined by the minimum condition. This argument depends only on the particular field content and the charge assignments and is valid also when higher order (in $\Lambda^{-1}$) terms are included. Given the superpotnetial $W_d$, eq. (\ref{wd}), the conditions of eq. (\ref{MinimumCond2}) can be solved to find the vacuum. A possible choice for the VEVs of  the relevant fields at the minimum reads
\bea
\langle \tilde{\xi}\rangle&=&0\nonumber\\
\langle \xi\rangle&=&u\nonumber\\
\langle \varphi^T\rangle&=&(v^T,0,0)\nonumber\\
\langle \varphi^S\rangle&=&(v^S,v^S,v^S)
\label{VEVS}
\eea
with $v^T=-3M/2g$, while $u$ and $(v^S)^2=-(g_4/3 g_3) u^2$ remain undetermined, parametrizing the degree of freedom along the flat direction (from now on we shall call $u$ this flat direction\footnote{Equivalently, in the basis
\begin{equation}
\begin{array}{cl}
u=&\left(\xi+i\sqrt{3g_3/g_4}(\varphi^S_1+\varphi^S_2+\varphi^S_3)\right)/4\\
u^\prime=&\left(\xi+i\sqrt{3g_3/g_4}(\varphi^S_1-\varphi^S_2-\varphi^S_3)\right)/4\\
u^{\prime\prime}=&\left(\xi-i\sqrt{3g_3/g_4}\,\,\varphi^S_1)\right/2\\
u^{\prime\prime\prime}=&\left(i\sqrt{3g_3/g_4}(\varphi^S_2-\varphi^S_3))\right/2,
\end{array}
\end{equation}
$u$ represents the flat direction, while $u^{\prime}=u^{\prime\prime}=u^{\prime\prime\prime}=0$.}). As explained above, in the limit of unbroken SUSY, the VEVs of the driving fields vanish exactly. An important feature of this model is that $A_4$ is broken by the VEVs of two different fields: $\varphi^T$, which leaves invariant a subgroup $Z_3\subset A_4$ and gives masses to the charged leptons, and $\varphi^S$, which breaks the $Z_3$ but leaves a $Z_2\subset A_4$ invariant and gives Majorana masses to the neutrinos. The $Z_3\subset A_4$ acts on the $A_4$ triplets as $T=\textrm{diag}(1,\omega,\omega^2)$, where $\omega=e^{i\frac{2\pi}{3}}$.

Supersimmetry breaking effects, when accounted for, contribute to the potential eq. (\ref{VSUSY}),
through soft masses for all scalars and A-type terms. Assuming that the main source of SUSY breaking preserves the flavor symmetry \cite{Feruglio:2009iu}, the main consequence of this breaking will be that small VEVs for the driving fields are generated, of order $\tilde{m}$. Furthermore a potential for the flat direction is also generated, removing the degeneracy of the minima $|\langle\varphi^S\rangle|\simeq\langle\xi\rangle=u$. Its potential can be written in the generic form
\begin{equation}\label{FDPot}
V(u)=\frac{a}{2}\tilde{m}^2|u|^2+\frac{b}{3}\frac{\tilde{m}^2}{\Lambda}u^3+h.c.+...,
\end{equation}
where the first term comes from SUSY breaking soft masses, while the second one is induced by non-renormalizable terms in the superpotential once the driving fields obtain VEVs of order $\tilde{m}$ (for instance the soft term $\tilde{m}(\varphi_0^T\varphi^S)\xi^2/\Lambda$ in the Lagrangian contributes to the cubic term once driving fields obtain their VEVs)\footnote{Renormalization effects are expected to drive the negative mass parameter $-|a|m^2$ to positive values at high energy or, equivalently, at large values of $u$, thus stabilizing the VEV of the flat direction at large values even in the absence of the term $\frac{b}{3}\frac{\tilde{m}^2}{\Lambda}u^3$.}. The coefficients $a$ and $b$ depend on the details of the model. We are interested in the case where the  parameter $a$ is negative. In this case, the true minimum of the potential is displaced from the origin $u=0$ and becomes very large, $u=(a/b)\Lambda$. This, together with the usual Higgs VEV $v_u=\langle h_u\rangle$, generates Majorana and Dirac mass matrices in the neutrino sector:
\be\label{Mmatrix}
m^D_\nu=y v_u \left(\begin{array}{ccc}
1& 0& 0\\
0& 0& 1\\
0& 1& 0
\end{array}\right),~~~~~~~~~~
M=\left(
\begin{array}{ccc}
A+2 B/3& -B/3& -B/3\\
-B/3& 2B/3& A-B/3\\
-B/3& A-B/3& 2 B/3
\end{array}
\right) u ~~~,
\ee
where $A\equiv 2 x_A$ and $B\equiv 2 x_B v^S/u$. Note that the Majorana mass for the RH neutrinos is proportional to the flat direction VEV $u$ and is in general very large. The heavy RH neutrinos can be integrated out and the mass matrix for the light neutrinos becomes $m_\nu=(m^D_\nu)^T M^{-1} m^D_\nu=-v_u^2y^2 M^{-1}$. Its eigenvalues are
\bea\label{lightnu}
m_{i}=\frac{y^2 v^2 \sin^2\beta}{M_{i}},
\label{Mn}
\eea
where $\tan\beta=v_u/v_d$ and $v=\sqrt{v_u^2+v_d^2}\approx 174\GeV$, while $v_{d}=\langle H_{d}\rangle$. The mixing matrix,
\begin{equation}
U^*M^{-1}U^{\dagger}=\textrm{diag}(M_1^{-1},M_3^{-1},M_3^{-1}),,
\end{equation}
with $M_i$ real and positive, has the tri-bimaximal form (up to a phase),
\begin{equation}
U^{\dagger}=U_{TB} U_{ph},\quad \textrm{with} \quad U_{TB}=\left(\begin{array}{ccc} 
\sqrt{2/3}&\sqrt{1/3}&0\\
-1/\sqrt{6}&1/\sqrt{3}&-1/\sqrt{2}\\
-1/\sqrt{6}&1/\sqrt{3}&1/\sqrt{2}
\end{array}\right),
\end{equation}
which is equivalent to eq. (\ref{TB}). Here $U_{ph}=\textrm{diag}(e^{i\alpha_1},e^{i\alpha_2},e^{i\alpha_3})$, encodes the phase information from the diagonalization of $M$. Small deviations from pure tri-bimaximal mixing can be accounted for by next-to-leading-order corrections in the expansion $1/\Lambda$.


\section{Symmetry Restoration and Breaking}
\label{SecThermal}

In this section we study the dynamics of the flavor symmetry breaking sector during the early universe. As explained above, the flavor symmetry is broken in this model in two steps: first at the scale $M$ by the VEV of $\varphi^T$, which leaves a subgroup $Z_3\subset A_4$ unbroken\footnote{For simplicity, we assume that the scale $M$ is bigger than the inflationary scale and that, consequently, $\varphi^T$ obtains a VEV before or during inflation. This assumption, however, has no influence on our conclusions.}, then, by the VEV of a supersymmetric flat direction which acquires a potential only in the presence of SUSY breaking terms, breaking $A_4$ completely. During inflation, the finite energy density of the vacuum breaks supersymmetry (the inflaton F- or D- components are necessary non-zero), inducing additional soft terms of order the Hubble constant $H_I$. In particular this generates a mass term
\begin{equation}\label{CorrInfl}
+\frac{c_I}{2}H^2_Iu^2.
\end{equation}
where the constant $c_I$ depends on the shape of the K\"ahler potential and superpotential. For the simplest case  of minimal couplings in the K\"ahler potential the coefficient is positive (for example, $c_I=3$ in F-term inflation) \cite{Dine:1995uk}. Although it is possible to arrange for negative values of $c_I$ (for instance, introducing non-minimal interactions between the inflaton and the flat direction in th K\"ahler potential, as done in Affleck-Dine bariogenesis \cite{Affleck:1984fy}), in the following we will assume it to be positive and of order unity. Consequently, during inflation, for $H_I\gg \tilde{m}$, the curvature of the potential is positive at the origin and the flat direction VEV is driven to zero: the $Z_3\subset A_4$ symmetry is unbroken and the RH neutrinos are massless, since they obtain their masses only through the last two terms of the superpotential, in eq. (\ref{wlss}).

After slow-roll, the inflaton reaches the minimum of its potential and begins to oscillate coherently around it, while beginning to decay. The universe, whose energy density is dominated by the oscillations, behaves as matter dominated and the Hubble parameter starts decreasing. Although most of the inflaton condensate will decay much later (depending on its decay width), a small fraction decays already during the first oscillations and quickly generates a (subdominant) thermal bath with temperature
\begin{equation}\label{Tosc}
T_{osc}\simeq k_{osc}\left(m_{pl}T_{RH}^2H\right)^{1/4},\quad\textrm{with}\quad k_{osc}=\left(\frac{9}{5\pi^3} \frac{g_*(T_{RH})}{g_*(T_{osc})}\right)^{1/8},
\end{equation}
where $g_*$ counts the number of effectively massless degrees of freedom at a given temperature and $T_{RH}$ defines the reheat temperature where hot big bang  cosmology begins. For the range of temperatures we are interested in, $k_{osc}\approx 0.4$. Note that $T_{osc}$ can generally be bigger than the reheat temperature. This fact, that a thermal bath is produced promptly after inflation (rather than there being a phase in which $H$ decreases but there is no radiation yet), gurantees a smooth transition between inflation and the subsequent reheating.

Eventually, after this phase, the inflaton decays completely and the universe remains filled with radiation. The constrains from gravitino over-production \cite{grav} tell us that the temperature at this stage (the reheat temperature, $T_{RH}$) cannot be bigger than about $10^{5\div 7} \GeV$. This, however, doesn't constrain the temperature $T_{osc}$ before inflaton decay, as the gravitinos produced at this stage are in fact diluted by the subsequent expansion\footnote{The power of four in the exponential of the relation $T_{osc}\propto H^{1/4}$ from eq. (\ref{Tosc}), ensures that at higher temperatures the expansion rate is much faster and that the gravitinos generated thermally before reheating at $T\sim T_{osc}$ are effectively diluted by the expansion and never exceed those produced around reheating, for $T\sim T_{RH}<T_{osc}$.}. 

So we see that ever since the end of slow-roll inflation, the universe contains a thermal bath of radiation with temperature $T_{osc}$ that decreases with the expansion. Since some of the particles in the thermal bath (the right-handed neutrinos, for instance) couple to the fields $\varphi^S$ and $\xi$ that make up the flat direction, they induce finite temperature corrections to its potential. For large values of $T$ and small values of $u$, the potential reads
\begin{equation}\label{Vt}
V_T(u)=\frac{1}{2}\left(-|a|\tilde{m}^2+c_TT^2\right)|u|^2+...,
\end{equation}
where dots stand for higher order terms in $u$. The positive coefficient $c_T$ receives contributions from all particles in the thermal bath that couple to the flat direction (driving fields and RH neutrinos) \cite{Dolan:1973qd}. For example the contribution coming from the RH neutrinos and sneutrinos is
\begin{equation}
c_T\supset \frac{1}{12}\left(18 x_A^2+\frac{20g_4}{3g_3}x_B^2\right),
\end{equation}
for simplicity, in what follows, we will assume it to be of order unity $c_T\sim{\cal O}(1)$.

As long as $T\gg\tilde{m}$, therefore, the flat direction acquires no expectation value, $u=0$ and the RH neutrinos remain massless. On the other hand, their Yukawa couplings are independent of the flat direction VEV, see eq. (\ref{wlss}), and mediate fast interactions between the thermal bath and the RH neutrinos (and sneutrinos), bringing them in thermal equilibrium. Their number density reads
\beq\label{n_RHN}
n_{N }^{eq}=\frac{2}{\pi^2}T^3.
\eeq
It is useful to express the RH neutrino abundance in terms of its ratio to the entropy density
\bea\label{YRHN}
Y_{N }^{eq}\equiv\frac{n_{N }^{eq}}{s}=\frac{45}{\pi^4 g_*},
\eea 
since this quantity has only a mild intrinsic temperature-dependence from $g_*$, counting the number of degrees of freedom in equilibrium at any given temperature.

The universe then continues its expansion and the temperature  decreases until it becomes comparable to the SUSY breaking scale, $T\sim \tilde{m}$. At this stage, the temperature contributions to the potential in eq. (\ref{Vt}) become smaller than the (negative) soft mass term and the curvature of the potential changes sign making the origin unstable: the flat direction is free to evolve to the true flavor-symmetry-breaking minimum of the potential. Note that for $T\gtrsim \tilde{m}$ the tunneling probability is way too suppressed \cite{Yamamoto:1985rd}, and the phase transition is most likely to take place classically (second order) as outlined so far, on a time scale set by the curvature of the potential, $\tau_{PT}\sim \tilde{m}^{-1}$.


\section{Leptogenesis}
\label{LG}

In \cite{Adhikary:2008au,Bertuzzo:2009im,Hagedorn:2009jy}, leptogenesis in the framework of models with a $A_4$ symmetry was studied and found that very high reheat temperatures, $T_{RH}\sim 5\times 10^{13}\GeV$  are required in order to produce the heavy RH neutrinos via scatterings in the thermal bath. This is in contrast with bounds coming from the overproduction of gravitinos in the supersymmetric context, as commented above. The symmetry restoration, however, drastically modifies this scenario. Its role for leptogenesis is  clear: during inflation, inflaton oscillations and reheating, the VEV of the flat direction responsible for the $Z_3\subset A_4$ symmetry breaking and for neutrino masses is kept in the origin by its interactions with other particles. At this stage the three RH neutrinos are massless and can be produced thermally even for very low reheat temperatures $T\gtrsim \tilde{m}$, independently of their would-be masses at $T=0$. When the temperature decreases below about $\tilde{m}$ (the exact temperature depending on the couplings of $\xi$ and $\varphi^S$ with the thermal bath), however, the phase transition takes place and the RH neutrinos become massive. Their masses, $M_i\gg \tilde{m}$, makes them completely out of equilibrium and eventually they decay via their CP-violating interactions producing an abundance of leptons over anti-leptons which is then transformed into the observed baryon asymmetry by sphalerons \cite{kuzmin}, as it is usually the case in Leptogenesis.

One can worry that the highly energetic decay products of the heavy RH neutrinos might overproduce gravitinos by thermal scattering. The decay products of the RH neutrinos, however are \emph{not} highly energetic. In fact, since the classical phase transition that gives the RH neutrinos a mass takes place over a finite time $\tau_{PT}\sim \tilde{m}^{-1}$, it is easy to show that the RH neutrinos decay before attaining their full masses. Indeed, their lifetime is
\beq\label{tauN}
\tau_{N_i}=\frac{8\pi}{[\hat{y}^{\dagger}\hat{y}]_{ii}M_{i}},
\eeq
which is inversely proportional to their mass $M_{i}\simeq u$, which increases during the phase transition ($\hat{y}$ denotes the Yukawa matrix in the basis where $M$ is diagonal). Hence, the RH neutrinos decay when $\tau_{N_i}\lesssim \tau_{PT}$ or, equivalently when
\beq\label{MiMin}
M_i\gtrsim M_i^{decay}= 8\pi\tilde{m}/[\hat{y}^{\dagger}\hat{y}]_{ii},
\eeq
producing a thermal bath with temperature
\beq
\tilde{T}=\frac{1}{\pi}\left(\frac{60}{g_*}\right)^{1/4}\left(\frac{M_i^{decay}}{\tilde{m}}\right)^{1/4}\tilde{m}.
\eeq
Using eq. (\ref{MiMin}), it's easy to see that for Yukawas bigger than $[\hat{y}^{\dagger}\hat{y}]_{ii}\gtrsim 10^{-13} $ this temperature is safely below the most conservative bounds for gravitino overproduction \cite{grav}, $T\lesssim 10^{5}\GeV$. Such low temperature also ensures that wash-out (and hence also flavor effects, which are normally present at temperatures below $10^{12}\GeV$,  \cite{Abada:2006ea}) are negligible, since $\Delta L=1$ inverse decays and $\Delta L=2$ processes are out of equilibrium for $\tilde{T}\ll M_i$, which is always the case.

Thus, the baryo-to-entropy ratio is given by
\bea\label{BAU}
Y_B\equiv \frac{n_B}{s}\approx \frac{8}{23}\epsilon Y_{N}^{eq},
\eea
where the approximate numerical factor comes  from the redistribution of lepton number to baryon number via sphaleron interactions, while the ratio of RH neutrinos number density to entropy at the time when RH neutrinos are still in equilibrium $Y_{N}^{eq}$ is given by eq. (\ref{YRHN}). The CP asymmetry $\epsilon$ vanishes in the limit of exact flavor symmetry \cite{Bertuzzo:2009im} and receives its first contributions from the terms
\begin{equation}\label{CPoperators}
\frac{\tilde{y}_1}{\Lambda}(t^n\,\varphi^T(l N)_s),\quad \textrm{and}\quad
\frac{\tilde{y}_2}{\Lambda}(t^n\,\varphi^T(l N)_a),
\end{equation}
where the subscripts $s,a$ specify in which way the flavor indices are contracted, see \cite{Altarelli:2005yx}. Again we can redefine the Yukawas $y_{1,2}=t^n\tilde{y}_{1,2}$ to absorb the factors of $t$; note that $y_{1,2}$ are of the same order as $y$ above. The CP asymmetries for the decay of $N_i$, using eqs.  (\ref{Mmatrix},\ref{CPoperators}) and refs. \cite{Bertuzzo:2009im,Covi:1996wh} are given by,
\begin{eqnarray}
\epsilon_1&=&\textrm{Re}(y_2)^2\frac{\eta^2}{8\pi}\left(\frac{8}{9}\,\frac{\textrm{Re}(y_1)^2}{\textrm{Re}(y_2)^2} \sin(\alpha_2-\alpha_1)f_{12}+\frac{1}{3} \sin(\alpha_3-\alpha_1)f_{13}\right)\nonumber\\
\epsilon_2&=&\textrm{Re}(y_2)^2\frac{\eta^2}{8\pi}\left(-\frac{8}{9}\,\frac{\textrm{Re}(y_1)^2}{\textrm{Re}(y_2)^2} \sin(\alpha_2-\alpha_1)f_{12}+\frac{2}{3} \sin(\alpha_3-\alpha_2)f_{23}\right)\\
\epsilon_3&=&\textrm{Re}(y_2)^2\frac{\eta^2}{8\pi}\left(-\frac{1}{3}\sin(\alpha_3-\alpha_1)f_{13}-\frac{2}{3}\sin(\alpha_3-\alpha_2)f_{23}\right)\nonumber.
\end{eqnarray}
The flavor symmetry breaking parameter\footnote{Note that, while $\varphi^S$ being a flat direction naturally obtains a large VEV even at low scales $\tilde{m}$, the VEV of $\varphi^T$ must be generated at a much higher scale $M\gg\tilde{m}$ in order to produce enough CP violation (and the right size of the $\tau$ Yukawa). The fact that $|\langle\varphi^T\rangle|\approx|\langle\varphi^S\rangle|$ is not surprising , as one might assume that the same physics that gives rise to the non-renormalizable terms in eq. (\ref{FDPot}), is responsible for the mass-term $M$ as happens, for example, in composite Higgs models \cite{CHM}, where naturally the Higgs vev is $v\lesssim \Lambda$.}
 $\eta=|\langle\varphi^T\rangle|/\Lambda\approx|\langle\varphi^S\rangle|/\Lambda\approx\langle \xi \rangle/\Lambda$ in realistic models is approximately \cite{Altarelli:2005yx}\beq\label{eta}
0.0022\lesssim\eta\lesssim 0.05,
\eeq
where the upper bound comes from the requirement that subleading corrections to the lepton mixing matrix not be too large, while the lower bound stems from the requirement that the tau Yukawa be small enough to justify a perturbative expansion. The function $f_{ij}$ depends on the ratios between RH neutrino masses as
\beq
f_{ij}=\frac{M_j}{M_i}\left(\frac{2-M_j^2/M_i^2}{1-M_j^2/M_i^2}-(1+M_j^2/M_i^2)\ln\left(\frac{1+M_j^2/M_i^2}{M_j^2/M_i^2}\right)\right),
\eeq 
and in the limit of strongly hierarchical RH neutrinos $M_j\gg M_i$ goes as $f_{ij}\sim (3/2) M_i/M_j$. In the models studied here, the masses $M_i$ and angles $\alpha_i$ are completely determined in terms of the solar and atmospheric mass parameters and the (unknown) mass of the lightest neutrino, $m_l$. 
\begin{figure}\label{fig}
\begin{minipage}[b]{0.5\linewidth} 
\centering
\includegraphics[height=5cm]{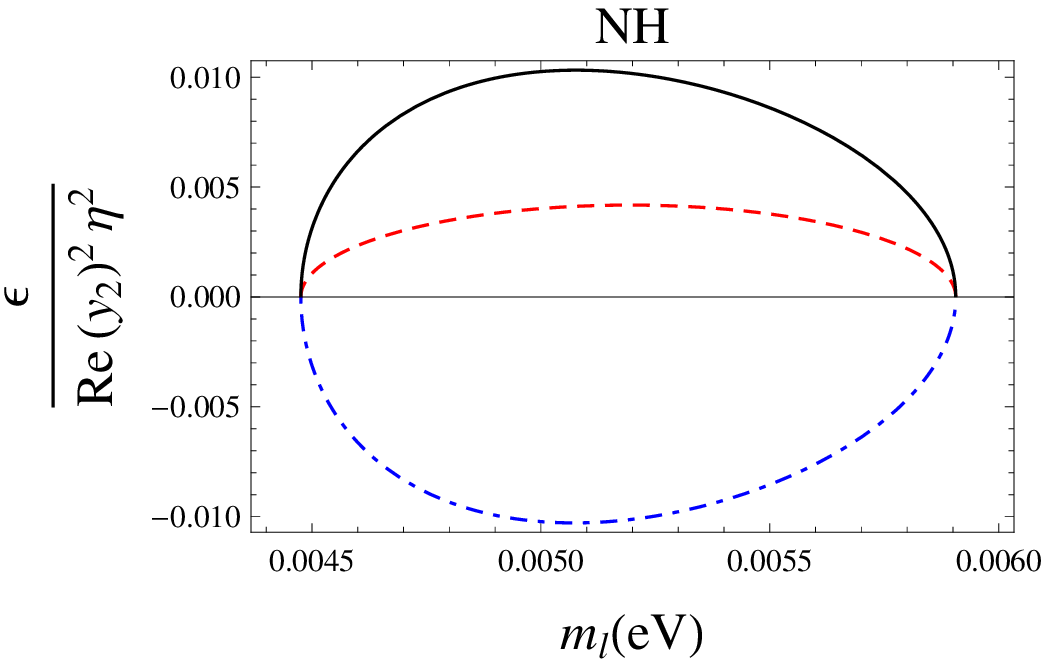}
\end{minipage}
\hspace{0.5cm} 
\begin{minipage}[b]{0.5\linewidth}
\centering
\includegraphics[height=5cm]{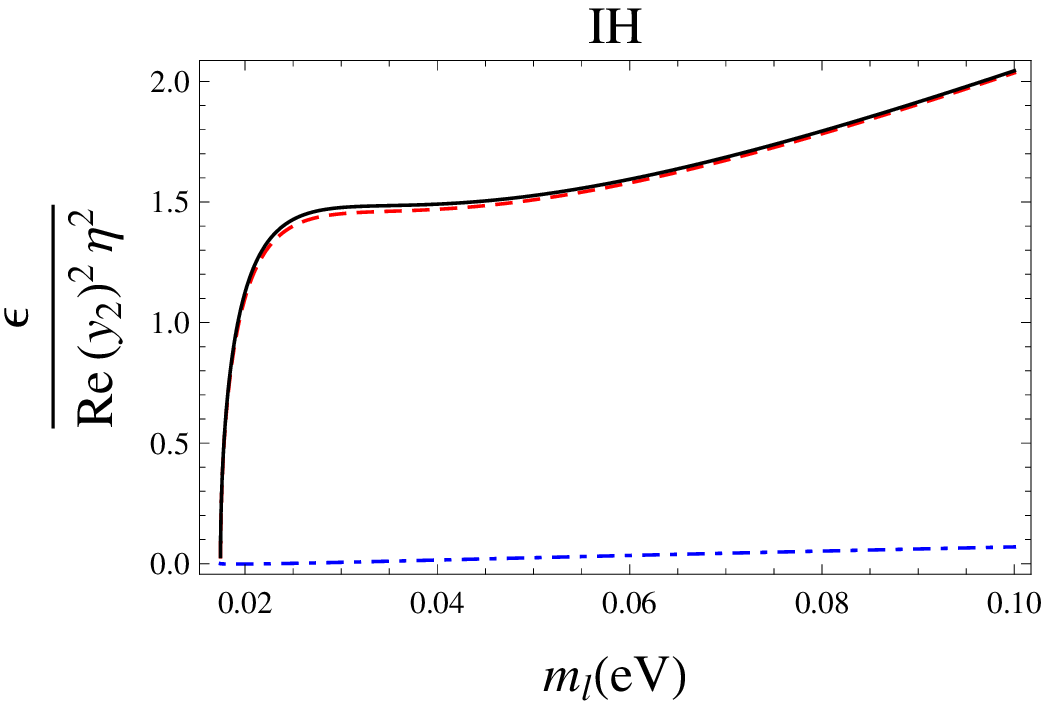}
\end{minipage}
\caption{\footnotesize The CP asymmetries $\epsilon_1$ (red, dashed), $\epsilon_2$ (black, solid) and $\epsilon_3$ (blue, dot-dashed), divided by $\eta^2 Re(y_2)^2$ for normal (left) and inverted (right) hierarchy as a function of the lightest neutrino mass, $m_{l}=m_1(m_3)$ for normal (inverted) hierarchy. We used $Re(y_2)^2=Re(y_1)^2$, while $\Delta m_{sol}^2=m_2^2-m_1^2=7.67\times10^{-5}\eV^2$ and $\Delta m_{atm}^2=|m_3^2-m_1^2|=2.46(2.37)\times10^{-3}\eV^2$, for NH (IH), see \cite{GonzalezGarcia:2007ib}.}
\end{figure}
Figure 1 shows that the largest\footnote{The three RH neutrinos obtain masses simultaneously and decay, almost at the same time. Therefore the lepton abundance will be produced mostly by the RH neutrino with the biggest $\epsilon$.} CP asymmetry  is $\epsilon_2$ and that,
\beq\label{epsilonOoM}
\frac{\epsilon_2}{\textrm{Re}(y_2)^2\eta^2}\lesssim\left\{ \begin{array}{ll}10^{-2}\,\,&\,\,\,\,\textrm{for NH}\\ 2\,\,&\,\,\,\,\textrm{for IH} \end{array}\right.,
\eeq
Here we have taken $\textrm{Re}(y_2)\approx\textrm{Re}(y_1)$ and NH (IH) stand for normal and inverted hierarchy respectively.  Comparing our results eqs. (\ref{BAU},\ref{epsilonOoM}) with the observed baryon asymmetry \cite{wmap} 
\beq
Y_B^{WMAP}=(0.87\pm0.03)\times 10^{-10},
\eeq
we find the following constraints on the Yukawa couplings,
\beq\label{LGSucc}
10^{-3}\lesssim y^2\approx y_1^2\approx y_2^2\lesssim1,
\eeq
for normal hierarchy, while
\beq\label{LGSucc2}
10^{-5}\lesssim y^2\approx y_1^2\approx y_2^2\lesssim10^{-2},
\eeq
for inverted hierarchy. The uncertainties depend on the value of $\eta$, eq. (\ref{eta}). In any case, this is largely compatible with the bounds given above for gravitino overproduction.


\section{Domain Walls}
\label{secDW}

Spontaneously broken discrete symmetries are commonly thought to be incompatible with standard cosmology, due to the existence of domain walls \cite{DW}. Domain walls are stable field configurations which interpolate between regions with different, but degenerate, vacuum states. As discussed above, spontaneously broken symmetries are generally restored at high temperatures and the phase transition occurs only when, during the expansion of the universe, the temperature is lowered below the critical value. The problem is that different regions of size $\sim H^{-1}$ are causally disconnected and, once the symmetry is broken, will find themselves in different vacua, divided by domain walls (the correlation length is reduced to $T^{-1}$ in the case of a second order phase transition). The energy density of domain walls is 
\beq\label{rhoDW}
\rho_{dw}\simeq \frac{\sigma}{R},
\eeq
where $R$ is the typical size of the domain walls, while $\sigma$ is the surface energy density. $R$ is proportional to the scale factor $r$ and the energy density of domain walls scales as $\rho_{dw}\sim r^{-1}$, much slower than radiation $\rho_{rad}\sim r^{-4}$ or matter, $\rho_{mat}\sim r^{-3}$. Therefore, if the phase transition takes place after inflation, these domain walls will eventually come to dominate the energy density of the universe leading to unacceptable observable signatures in the cosmic microwave background. On the other hand, if the scale of the phase transition is higher than the inflationary scale, then one can assume that the universe expanded enough that the relevant portion around the present horizon is clear from domain walls. This is the case for the $\varphi^T$ phase transition, assuming that the scale $M$ in eq. (\ref{wd}) is larger than the inflationary scale. Unfortunately, $\varphi^S$ obtains a VEV at $T\sim \tilde{m}$ well after inflation, and the domain walls produced at this stage, separating vacua with different $Z_3\subset A_4$ charge, can lead to cosmological catastrophe. In this section we discuss three different solutions to this problem.

\noindent
{\bf 1.} The discussion above assumes that the symmetry $A_4$ is exact. But it's not. When the model is extended to include quarks \cite{FeruglioQuarks}, these transform under a chiral representation which is anomalous: the $A_4$ symmetry is only an apparent symmetry of the classical action, broken at the quantum level. The same mechanism that gives the axion a mass, here lifts the degeneracy between vacua and leads to the disappearance of the domain walls \cite{Preskill:1991kd}.

Referring to \cite{FeruglioQuarks} for the details of how the $A_4$ symmetry can be extended to quarks, we recall here how quarks transform under the $Z_3\subset A_4$,
\beq
\begin{array}{l}
u_R\rightarrow w^2\, u_R\\
d_R\rightarrow w^2 \,d_R\\
q_L\rightarrow w \,q_L,
\end{array}
\eeq
with $w=e^{i2\pi/3}$, while the other quarks are singlets. This transformation is chiral and it has a QCD anomaly, in the sense of ref. \cite{Preskill:1991kd} - it acts non-trivially on the quark mass matrix $M_q$. Since a phase shift in $\arg \det(M_q)$ corresponds to a shift in the $\theta_{QCD}$-angle,  we see that the action of $Z_3\subset A_4$ interpolates between $\theta_{QCD}=0,2\pi/3,4\pi/3$. Now the same physics that gives the axion a mass, lifts the original degeneracy by a small amount $\Delta \rho$, proportional to the axion mass,
\beq
\Delta \rho \simeq f_{\pi}^2m_{\pi}^2\frac{m_u}{m_u+m_d}
\eeq
at $T=0$, where $m_{\pi}$, $f_{\pi}$ are the pion mass and decay constant and $m_{u,d}$ are the $u,d$-quark masses. At large temperatures, this small energy splitting has no influence on the dynamics of domain walls, but at temperatures below the chiral phase transition, $T\lesssim T_{QCD}=250 \MeV$, the pressure caused by this energy difference will become sizable and the domain walls collapse, \cite{Preskill:1991kd}. We need to make sure that this collapse takes place before the domain walls dominate the energy density of the universe, which would otherwise lead to power-law inflation that empties the universe from matter and radiation.

In our model the domain walls have width $\delta\sim \tilde{m}^{-1}$ and false vacuum energy $\rho_v\sim\tilde{m}^2u^2$, resulting \cite{DW}  in a surface energy density $\sigma=\delta\rho_v\sim\tilde{m}u^2$. The tension, with a force per unit area $F_t\sim \sigma/R$, tends to drive small walls to collapse, while wiping away inhomogeneities on larger walls. On the other hand, friction $F_f\sim-T^4 v$ ($v$ being the wall velocity with respect to the bath with temperature $T$), caused by particles changing their masses while crossing the wall, retards this collapse and only scales much smaller than the horizon are wiped away. The size of the typical scales surviving is set by a balance between tension and friction,
\begin{equation}
\frac{\sigma}{R}=T^4 v.
\end{equation}
Indeed, the time required to straighten a wall of size R is
\begin{equation}
t_d=\frac{R}{v}\simeq \frac{R^2 T^4}{\sigma},
\end{equation}
which means that in one Hubble time, $t_d=H^{-1}\simeq M_{pl}/(g_*^{1/2}T^2)$, only scales bigger than
\begin{equation}
R_s\simeq\frac{\sqrt{\sigma M_{pl}}}{g_*^{1/4}T^3}
\end{equation}
will remain. This sets the typical size of domain walls at temperature $T$. Using eq. (\ref{rhoDW}) we can compare the energy stored in the domain walls at the QCD phase transition with the energy density of the universe at temperature  $T_{QCD}$:
\begin{equation}
\frac{\rho_{dw}}{\rho_{rad}}\simeq \frac{1}{\pi g_*^{3/4}}\sqrt{\frac{\tilde{m}}{M_{pl}}}\frac{u}{T_{QCD}}\sim \frac{u}{10^{10} \GeV}
\end{equation}
In order for the domain walls never to dominate the energy density of the universe, we need that the flat direction VEV,
\beq\label{umax}
u \lesssim 10^{10} \GeV.
\eeq
Recall from eq. (\ref{Mmatrix}) that the flat direction VEV is of order the RH neutrino masses, $M_i\sim u$, and that these are fixed through eq. (\ref{lightnu}) by the size of the Yukawas $y$ and the light neutrino masses. Thus, the bound of eq. (\ref{umax}) can be thought in our model as a bound on the Yukawa couplings,
\beq
\hat{y}^\dagger\hat{y}\lesssim 10^{-5},
\eeq
where we have used $m_{l}=m_{atm}=0.05 \eV$. Comparing this with the results of eq. (\ref{LGSucc}), shows that this solution of the domain wall problem is compatible with leptogenesis only in the IH case, while the NH case is disfavored since there domain walls would come to dominate the energy density of the universe before their collapse. The fields $\varphi^S$ and $\xi$ along the flat direction also carry a non vanishing $Z_3$ charge which is liable to yield to its own domain walls. This symmetry however is violated explicitly by the $\mu$-term of SUSY, $\mu H_u H_d$. Due to this relatively large explicit violation, similarly to the mechanism just showed, the domain walls will collapse long before the QCD phase transition\footnote{Alternatively, the $Z_3$ symmetry - its only role being the separation between charged and neutral lepton sectors - can be substituted by a larger discrete symmetry (such as $Z_4$), with anomalous charge assignments. In this case its domain walls will disappear together with the ones from the broken $Z_3\subset A_4$ symmetry.}.                                                                                                                                                                                                                                                                                                                                                                                                                                                                                                                        

\noindent
{\bf 2.} It must be noted that the mechanism explained above relies on the fact that the discrete symmetry $A_4$ is not violated by the gravitational interactions, which is true, for instance if the symmetry originates within a continuous gauge group. However, if a small, Planck-suppressed symmetry breaking term is present in the effective Lagrangian, the situation is different. Consider for example the term
\begin{equation}\label{explicit}
\frac{1}{M_{pl}}(\varphi^S\varphi^S)^\prime(\varphi^S\varphi^S)\xi_0,
\end{equation}
which explicitly breaks $Z_3\subset A_4$, since the contraction $(...)^\prime$ is not invariant. The inclusions of these terms does not compromise the vacuum alignment, since they are suppressed by at least a factor $\Lambda/M_{pl}$ with respect to symmetry preserving next-to-leading-order operators, which already give contributions of order $\eta$, \cite{Altarelli:2005yx}.  When $\varphi^S$ and $\xi_0$ acquire VEVs, the contribution from this term would be enough to erase completely the barrier separating different vacua. In this case, independently on the value of $u$, the domain walls will not even form. Indeed, the energy difference between vacua provided by the term eq. (\ref{explicit}) is of order $\tilde{m}\,u^4/M_{pl}$, and for $u\gtrsim10^{10}\GeV$ is bigger then the height of the barrier between vacua, which is of order $\tilde{m}^2u^2/3$.

{\noindent}
{\bf 3.} Before concluding, we note that in more conventional models of leptogenesis (or where the barion asymmetry is produced through another mechanism, such as \cite{Giudice:2008gu} or Affleck-Dine bariogenesis \cite{Affleck:1984fy}), the domain wall problem can be solved through a non-restoration of the discrete symmetry at high temperatures. Indeed, it is possible to arrange for (non-minimal)  couplings in the K\"ahler potential between the inflaton and the fields making up the flat direction ($\xi$ and $\varphi^S$), such that the term of eq. (\ref{CorrInfl}) setting the curvature of the flat direction during inflation, is negative \cite{Dine:1995uk}. In this case a phase transition takes place during inflation and the domain walls produced are effectively diluted by the subsequent expansion. Normally the symmetry would be restored during the reheating phase, reintroducing the domain wall problem. However, since flat directions acquire very large VEVs in comparison to their curvature, they induce very large masses to all fields which couple to them in a renormalizable way. These heavy fields decouple from the effective theory and are no longer able to mediate temperature effects to the flat direction, so that its potential is approximately given by the $T=0$ potential of eq. (\ref{FDPot}). In this case the symmetry is never restored at high temperatures and domain walls do not cause any problem.

To summarize, if the condition eq. (\ref{umax}) is met, then the domain walls collapse at the QCD phase transition without consequences. Otherwise, a small gravity mediated explicit breaking of the $A_4$ symmetry is needed. Alternatively, if the baryon asymmetry is produced by a different mechanism than the one discussed in the first part of this article, it might be possible to arrange for the symmetry to be broken already during inflation: in this case the symmetry is never restored and domain walls do not form.


\section{Conclusions}
We have analyzed thermal leptogenesis in the framework of models with an $A_4$ flavor symmetry that naturally reproduce tri-bimaximal neutrino mixing, although our findings apply to the more general case where the RH neutrino masses arise from a mechanism of spontaneous symmetry breaking. We found that leptogenesis can be successful, without any fine-tuning, even at low energies well below the gravitino bound, independently of the masses of RH neutrinos. This is possible thank to a mechanism of symmetry restoration, already present in the $A_4$-based models, that ensures that the RH neutrino masses vanish at temperatures $T\gtrsim 1 \TeV$. Hence, during the early universe, the RH neutrinos can be produced copiously by thermal scattering, without requiring very high temperatures that would lead to the overproduction of gravitinos, in contrast with the standard cosmological scenario. The symmetry breaking discussed here takes place along a supersymmetric flat direction, lifted only by SUSY breaking effects. Its potential is therefore naturally almost flat and induces very large field VEVs, despite its scale being of order the soft SUSY breaking masses, $\tilde{m}\sim {\cal O}(10^2 - 10^3) \GeV$. Hence the phase transition that gives a mass to the RH neutrinos takes place only at temperatures much smaller than the would-be RH neutrino masses.
Eventually the heavy neutrinos decay out of equilibrium via their CP-violating interactions producing an abundance of leptons over anti-leptons which is then transformed into the observed baryon asymmetry by sphaleron interactions.

The $A_4$-based models are very predictive in terms of the neutrino mixing angles. Indeed all 3 mixing angles are predicted to depart from their pure tri-bimaximal values, eq. (\ref{TB}) by the same amount. Since measured deviation of $\theta_{12}$ are very small, see eq. (\ref{data}), the same must be true for $\theta_{13}$. So, observation of $\theta_{13}\neq 0$ in future experiments will be able to exclude models with tri-bimaximal mixing in the neutrino sector.

Models with discrete symmetry spontaneously broken at energies below the inflationary scale, generally suffer from a domain wall problem. We showed that the $A_4$-based models, when extended to account for quark masses, posses a QCD anomaly, in the sense that the $A_4$ acts non-trivially on the quark mass matrix. In this case, the same mechanism that gives a mass to the axion, here lifts the degeneracy between vacua on opposite sides of the domain walls. We found that for neutrinos with inverted hierarchy this solution of the domain wall problem is compatible with our scenario of leptogenesis. For neutrinos with a normal hierarchy, on the contrary, this effect is not enough to avoid a domain wall dominated universe if we insist on the mechanism of leptogenesis as described above. In both cases, if the discrete symmetries are broken explicitly by the gravitational interactions, there remains no barrier separating different vacua and domain walls never form.

\vspace{2cm}
\centerline{\bf Ackowledgements}
\noindent
I'm particularly indebted with Ferruccio Feruglio for several useful discussions and with Antonio Riotto for his comments. I also thank Ben Gripaios and Luca Merlo for interesting conversations. This work was supported by the Fondazione Cariparo Excellence Grant \emph{LHCosmo} and in part by the European Programme \emph{Unification in the LHC Era}, contract PITN-GA-2009-237920 (UNILHC), and the Research and Training Network UniverseNet.


\begin{thebibliography}{99}

\bibitem{LG}
M.~Fukugita and T.~Yanagida,
  Phys.\ Lett.\ B {\bf 174}, 45 (1986). For a Review, see
  W.~Buchmuller, R.~D.~Peccei and T.~Yanagida,
  Ann.\ Rev.\ Nucl.\ Part.\ Sci.\  {\bf 55}, 311 (2005); or 
S.~Davidson, E.~Nardi and Y.~Nir.
  
 \bibitem{seesaw} 
M. Gell-Mann, P. Ramond and R. Slansky, 
in {\it Supergravity}, edited by P. van Nieuwenhuizen and D. Freedman, 
(North-Holland, 1979), p.~315; 
S.L. Glashow, in Quarks and Leptons, Carg\`ese, eds. M. L\'evy et al., 
(Plenum, 1980, New-York), p. 707; 
T. Yanagida, in {\it Proceedings of the Workshop on the Unified Theory 
and the Baryon Number in the Universe}, edited by O. Sawada and
A. Sugamoto (KEK Report No.~79-18, Tsukuba, 1979), p.~95; 
R.N.~Mohapatra and G. Senjanovi\'{c}, Phys. Rev. Lett. {\bf 44}, (1980) 912.

\bibitem{Froggatt:1978nt}
  C.~D.~Froggatt and H.~B.~Nielsen,
  Nucl.\ Phys.\  B {\bf 147} (1979) 277.

\bibitem{Altarelli:2005yp}
For a Review, see:  G.~Altarelli and F.~Feruglio,
  arXiv:1002.0211 [hep-ph].


\bibitem{Altarelli:2005yx}
  G.~Altarelli and F.~Feruglio,
  Nucl.\ Phys.\  B {\bf 741} (2006) 215.
  
\bibitem{Basboll:2007vt}
  A.~Basboll, D.~Maybury, F.~Riva and S.~M.~West,
  Phys.\ Rev.\  D {\bf 76}, 065005 (2007);
  A.~Riotto and F.~Riva,
  Phys.\ Lett.\  B {\bf 670}, 169 (2008).
  
  \bibitem{Giudice:2008gu}
  G.~F.~Giudice, L.~Mether, A.~Riotto and F.~Riva,
  Phys.\ Lett.\  B {\bf 664}, 21 (2008).

\bibitem{Affleck:1984fy}
  I.~Affleck and M.~Dine,
  Nucl.\ Phys.\  B {\bf 249} (1985) 361.
  
\bibitem{Dine:1995uk}
  M.~Dine, L.~Randall and S.~D.~Thomas,
  Phys.\ Rev.\ Lett.\  {\bf 75}, 398 (1995).
  
  \bibitem{kuzmin} 
  V.~A.~Kuzmin, V.~A.~Rubakov and M.~E.~Shaposhnikov,
  Phys.\ Lett.\  B {\bf 155}, 36 (1985).

\bibitem{DW}
  Y.~B.~Zeldovich, I.~Y.~Kobzarev and L.~B.~Okun,
  Zh.\ Eksp.\ Teor.\ Fiz.\  {\bf 67} (1974) 3
  [Sov.\ Phys.\ JETP {\bf 40} (1974) 1]; 
  T.~W.~B.~Kibble,
  J.\ Phys.\ A  {\bf 9}, 1387 (1976); 
  A.~Vilenkin,
  Phys.\ Rept.\  {\bf 121} (1985) 263.

\bibitem{Schwetz:2008er}
  T.~Schwetz, M.~A.~Tortola and J.~W.~F.~Valle,
  New J.\ Phys.\  {\bf 10} (2008) 113011.

\bibitem{Harrison:2002er}
  P.~F.~Harrison, D.~H.~Perkins and W.~G.~Scott,
  Phys.\ Lett.\  B {\bf 530} (2002) 167.
  
\bibitem{Feruglio:2009iu}
  F.~Feruglio, C.~Hagedorn and L.~Merlo,
  arXiv:0910.4058 [hep-ph].
 
\bibitem{grav} M.~Y.~Khlopov and A.~D.~Linde,
  Phys.\ Lett.\  B {\bf 138}, 265 (1984); 
J.~R.~Ellis, J.~E.~Kim and D.~V.~Nanopoulos,
  Phys.\ Lett.\  B {\bf 145}, 181 (1984); 
M.~Kawasaki, K.~Kohri and T.~Moroi,
  Phys.\ Lett.\  B {\bf 625}, 7 (2005). See also:
    I.~V.~Falomkin, G.~B.~Pontecorvo, M.~G.~Sapozhnikov, M.~Y.~Khlopov, F.~Balestra and G.~Piragino,
  Nuovo Cim.\  A {\bf 79} (1984) 193
  [Yad.\ Fiz.\  {\bf 39} (1984) 990].
  
  
\bibitem{Dolan:1973qd}
  L.~Dolan and R.~Jackiw,
  Phys.\ Rev.\  D {\bf 9} (1974) 3320.
  
\bibitem{Yamamoto:1985rd}
  K.~Yamamoto,
  Phys.\ Lett.\  B {\bf 168} (1986) 341.
  
    \bibitem{Adhikary:2008au}
 B.~Adhikary and A.~Ghosal,
 Phys.\ Rev.\  D {\bf 78}, 073007 (2008).
 
  
\bibitem{Bertuzzo:2009im}
  E.~Bertuzzo, P.~Di Bari, F.~Feruglio and E.~Nardi,
  JHEP {\bf 0911} (2009) 036.
 
  \bibitem{Hagedorn:2009jy}
  C.~Hagedorn, E.~Molinaro and S.~T.~Petcov,
  JHEP {\bf 0909} (2009) 115.
  
\bibitem{Abada:2006ea}
  A.~Abada, S.~Davidson, A.~Ibarra, F.~X.~Josse-Michaux, M.~Losada and A.~Riotto,
  JHEP {\bf 0609} (2006) 010;
  A.~Abada, S.~Davidson, F.~X.~Josse-Michaux, M.~Losada and A.~Riotto,
  JCAP {\bf 0604} (2006) 004.
  
\bibitem{Covi:1996wh}
  L.~Covi, E.~Roulet and F.~Vissani,
  Phys.\ Lett.\  B {\bf 384} (1996) 169.
  
  
  \bibitem{CHM}
    K.~Agashe, R.~Contino and A.~Pomarol,
  Nucl.\ Phys.\  B {\bf 719}, 165 (2005)
; see also
   B.~Gripaios, A.~Pomarol, F.~Riva and J.~Serra,
  JHEP {\bf 0904} (2009) 070.
  
\bibitem{GonzalezGarcia:2007ib}
  M.~C.~Gonzalez-Garcia and M.~Maltoni,
  Phys.\ Rept.\  {\bf 460} (2008) 1.
  
  \bibitem{wmap} 
 J.~Dunkley {\it et al.}  [WMAP Collaboration],
    arXiv:0803.0586 [astro-ph].  
    
\bibitem{FeruglioQuarks}
  F.~Feruglio, C.~Hagedorn, Y.~Lin and L.~Merlo,
  Nucl.\ Phys.\  B {\bf 775} (2007) 120.

\bibitem{Preskill:1991kd}
  J.~Preskill, S.~P.~Trivedi, F.~Wilczek and M.~B.~Wise,
  Nucl.\ Phys.\  B {\bf 363} (1991) 207.


\end{thebibliography}
\end{document}